# Characterization of the Torsional Piezoelectric-like Response of Tantalum Trisulfide Associated with Charge-Density-Wave Depinning

J. Nichols, D. Dominko, L. Ladino, J. Zhou, and J.W. Brill

Department of Physics and Astronomy
University of Kentucky
Lexington, KY 40506-0055

**ABSTRACT**

We have studied the frequency and voltage dependence of voltage-induced torsional strains in orthorhombic TaS$_3$ [V. Ya. Pokrovskii, *et al*, Phys. Rev. Lett. **98**, 206404 (2007)] by measuring the modulation of the resonant frequency of an RF cavity containing the sample. The strain has an onset voltage below the charge-density-wave (CDW) threshold voltages associated with changes in shear compliance and resistance, suggesting that the strain is associated with polarization of the CDW rather than CDW current. Measurements with square-wave voltages show that the strain is very sluggish, not even reaching its dc value at a frequency of 0.1 Hz, but the dynamics appear to be very sample dependent. By applying oscillating torque while biasing the sample with a dc current, we have also looked for strain induced voltage in the sample; none is observed at the low biases where the voltage-induced strains first occur, but an induced voltage is observed at higher biases, probably associated with strain-dependent CDW conductance.

Quasi-one dimensional-conductors with sliding charge-density-waves (CDWs) are well known for their unusual, nonlinear electronic properties associated with polarization and motion of the CDW above a depinning threshold voltage, $V_T$ [1]. CDW depinning is also accompanied by changes in mechanical properties [2]; the largest observed change is for the low-frequency shear compliance of the orthorhombic polytype of tantalum trisulfide (o-TaS$_3$), which increases by ~ 25% with depinning [2,3,4]. This softening was associated with the crystal strain dependence of the CDW wavevector causing relaxational changes in the domain configuration of the CDW [4,5]. In 2007, Pokrovskii *et al* reported that when a voltage near threshold is applied to a crystal of o-TaS$_3$ which is free to distort (i.e. has one end mechanically clamped and one end mechanically free), the crystal twists by ~ 1° [6,7]. The twist direction reverses for a voltage of the opposite polarity, so that the angle is a hysteretic function of dc voltage. While this twist was only observed in the CDW state, each crystal always had a well-defined direction of twist (for each voltage polarity), even if cycled through its CDW phase transition and even if cut in half, suggesting that this voltage-induced torsional strain (VITS) was associated with the interaction of CDW current or strain with a lattice defect which extends most of the length of the sample [6]. (Note that no chiral features have ever been reported in either the lattice or CDW structure of o-TaS$_3$ [6] and, as discussed in [6], even an undetected screw axis could not explain the observed VITS.)

More recently, Pokrovskii *et al* found that, by applying square-wave voltages to the sample, the hysteretic VITS was very sluggish, disappearing for applied frequencies near 100 Hz. However, they also observed an additional small (< 0.01°) VITS at high frequencies which, while sensitive to the presence of the CDW, was not sensitive to depinning [8]. Similar torsional effects were also observed in other CDW conductors [8]. Finally, they found that if an o-TaS$_3$ crystal, when biased well above threshold with a dc current, was twisted with a high frequency (> 6 kHz) torque, a small ac voltage was induced; i.e. o-TaS$_3$ acted as "torsional piezo-resistor" [9].

In this paper, we report on details of the voltage and frequency dependence of the "sluggish" component of this torsional piezoelectric-like strain in o-TaS$_3$. (Our experimental techniques do not have the sensitivity to observe the small "fast twists".) We compare these dependences with those of changes in the shear compliance for the same samples. We also studied the voltage dependence of the piezo-resistance. While piezo-resistance was in fact also observed near threshold, its voltage dependence and fast dynamic response suggest that it is not the "direct piezoelectric" effect, i.e. strain induced voltage, that might be expected to complement the VITS response.

Extensive measurements were done on two o-TaS$_3$ crystals at T = 78 K. Both ribbon shaped crystals were ~ 2 mm long and ~ 10 μm wide. Crystal A was ~ 4 μm thick while crystal B was < 2 μm thick. One end of each o-TaS$_3$ crystal was glued with conducting paint to a rigid voltage contact while the other end was glued to a small magnetized steel wire (25 μm diameter, ~ 3 mm long, perpendicular to the sample). The wire was also glued to a NbSe$_3$ crystal, with dimensions comparable to the o-TaS$_3$, which was glued to the second voltage contact (Fig. 2b inset). Since NbSe$_3$ remains metallic while o-TaS$_3$ becomes semiconducting below their CDW transitions [1], the voltage drop across the NbSe$_3$ is negligible at low temperatures. However, because the shear moduli of the two materials are presumably comparable, the NbSe$_3$ crystal acts as a torsional spring in parallel with the sample, reducing both the voltage-induced torsional strain and measured changes in shear compliance (by 30% for sample A and 80% for B).

The samples were mounted in an RF helical resonator cavity [10] as shown in Fig. 2b inset, with the tip of the helix ~ 0.2 mm from the end of the steel wire, so that when the sample twisted it changed the resonant frequency of the cavity. "Static" twists (as functions of dc voltage applied to the sample) were measured (in arbitrary units) by driving the cavity with a frequency modulated carrier [3]; resulting measurements were very susceptible to drifts in the electronics. More sensitive measurements of strain, also in arbitrary units, were made by applying square-wave voltages to the sample and measuring the modulated response, at the square-wave frequency, of the cavity driven at its resonance. Relative changes in the effective shear compliance (i.e. including its elastic loading by the NbSe$_3$ "spring") were measured by applying torque to the magnetized wire through an oscillating magnetic field and measuring the modulated response of the cavity [3,4]. Similarly, any ac voltage induced in the sample induced by the twist could be measured.

Typical "static" torsional strain vs. dc voltage hysteresis loops for the two samples are shown in Fig. 1a, with typical cycle times of ~ 1 hr. While the responses of the two samples appear quite similar, it should be emphasized that the arbitrary units for each are not the same; the FM responses for the two samples have been normalized to give hysteresis loops of roughly the same widths. (From comparison with the results of Pokrovskii *et al* [6], we estimate that "1 arb. unit" ~ 1 degree.) Figure 1b shows the dc voltage dependence of the relative changes of the effective shear compliance, J, measured with 10 Hz magnetic torque, and dc resistance for the two samples. (These were measured with decreasing positive current in the sample. The resistance values below threshold are, in fact, hysteretic [6], where the hysteresis is associated with the CDW polarization [1,11]. The resistances and torsional resonant frequencies were 57 k$\Omega$ and over 100 Hz for crystal A and 134 k$\Omega$ and ~ 30 Hz for crystal B.)

As shown in Figure 1a, the VITS hysteresis loops were slightly asymmetric for these samples. For increasing positive voltages, both samples start twisting at $V_{on}$ ~ 20 mV with the twist saturating at $V_{sat}$ ~ 80 mV, while for increasing negative voltages, $V_{on}$ ~ -40 mV and $V_{sat}$ ~ -100 mV. The VITS response measured for other samples was more symmetric, as shown in the inset to Fig.1a. The threshold voltages determined from the changes in compliance was $V_T$ ~ ± 50 mV for both samples, below the voltages at which the resistance starts dropping rapidly [4]. Therefore, the elastic threshold lies between $V_{on}$ and $V_{sat}$. In samples for which voltage contacts strongly pin the CDW, the CDW can become depinned in the bulk of the sample, so that its local phase strains (i.e. the CDW polarizes) at voltages below that at which it becomes depinned at the contacts, which allows dc CDW current to flow [11], and we have previously observed that the difference in these two voltages, the "phase-slip voltage", could be as large as 60 mV for o-TaS$_3$ samples [12]. This suggests that the VITS effect may be associated with strains of the CDW, rather than CDW current, interacting with crystal defects.

As mentioned above, the VITS can be measured more cleanly by applying oscillating voltages to the sample and measuring the resulting torsional oscillations. We did this by applying symmetric square-wave voltages (±V) of varying amplitude and frequency; for voltages above $V_{on}$, we expect to start swinging the sample through the hysteresis loop. The results for the two samples are shown in Figure 2, where the responses both in-phase and in quadrature with the applied square wave are shown. (As before, the response is shown in arbitrary units, chosen to approximately match those of Figure 1a, i.e. 1 arb. unit ~ 1°.)

For both samples, the onset for the VITS is ~ 20 mV ~ $+V_{on}$, as expected. The onset is independent of frequency, as shown more clearly for sample A in Fig. 2a inset, where the magnitude of the VITS signal is plotted. However, our initial 10 Hz measurements on A were surprising in that the VITS signal did not saturate for voltages < 250 mV, well-beyond the value of $V_{sat}$. As shown in Fig. 2a, this is in fact a dynamic effect. The large quadrature signal at 250 mV and 10 Hz shows that the VITS signal lags the applied square wave considerably (by ~ 15°). With decreasing frequency, the high voltage in-phase response increases and saturates at lower voltages, while the quadrature signal decreases. However, even at 0.1 Hz, the saturation voltage ~ 130 mV is greater than the dc value of $V_{sat}$.

It is interesting to compare the dynamics of the VITS effect with that of the change in the shear compliance, which is associated with relaxation of the CDW domain configuration as the sample is twisted. In the inset to Fig 1b, we compare the phase shifts, at 10 and 3 Hz, of the change in compliance, i.e. $\tan^{-1}[\Delta J(quadrature)/\Delta J(in-phase)]$ with that of the VITS for sample A. It is seen that the VITS phase shift is several times larger than that of the compliance change, indicating the torsional effect is much more sluggish.

Despite their very similar hysteresis loops, the voltage and frequency dependences of the square-wave VITS for sample B, shown in Fig. 2b, are very different from those of sample A. For B, the quadrature response has nearly symmetric peaks at the voltage at which the in-phase response grows, suggestive of a voltage dependent relaxation time (with $\tau \sim V^{-6}$, a much stronger dependence than observed for the compliance change [4] or polarization [12]). However, the in-phase response decreases with increasing voltage at the highest voltages and the quadrature response becomes negative, indicating that the VITS signal overshoots with voltage reversal and then decays slightly with time. The different dynamic responses of the two samples presumably reflect different spatial distributions of lattice defects interacting with the CDW polarization.

In particular, the CDW polarization corresponds to an increase in wavelength near one current contact and decrease near the other [11-13]. While a common simplifying assumption [11-13] is that the CDW wave fronts remain normal to the crystal axis, surface defects have been observed to cause the wave fronts to bend [14]. The VITS effect suggests that defects can also cause the wave fronts to become chiral (with opposite chiralities at the two ends), and these CDW strains then put torsional stress on the sample [15]. Near threshold, the longitudinal CDW strains diffuse through the sample [16], and for a 2 mm long sample at T = 80 K would have a time constant ~ 5 ms [12]. The slower VITS signals then indicate that, hindered by their interaction with lattice defects, the torsional CDW strains change more slowly. The dynamic response will be the subject of future study.

The piezoelectric-like torsional response to applied voltage suggests that the complimentary effect, i.e. a twist-induced EMF, should also be observed near threshold, although the expected magnitude of the voltage is not clear, in view of the nonlinear and hysteretic behavior of the VITS as well as the relatively high conductance of the sample. We looked for a twist-induced EMF by oscillating the sample with an ac magnetic field while applying a dc bias current. Typical results for the two samples for a 10 Hz magnetic field, showing the induced EMF's in-phase and in quadrature with the torque, are shown in Figure 3. For the results shown, the strains were about 5 times larger than the VITS. The induced EMF was observed to be proportional to the magnitude of the oscillating magnetic field, so that for a field that gave a strain comparable to the VITS, the induced emf was comparable to the noise.

It is striking that the induced EMF starts growing (for both samples) at ~ 100 mV, i.e. above $V_{on}$. Furthermore, the induced voltage seems to be much faster than the VITS, as shown by its negligible quadrature component and the fact that the induced voltages at 3 Hz (not shown) were the same as those at 10 Hz. This suggests that the observed EMF is not complimentary to the VITS; i.e. any complimentary EMF is below our noise level (~ 3 µV). Instead, we suggest that our signal has the same origin as the high frequency "torsional piezo-resistance" reported in Ref. [9]. In fact, since the induced EMF starts growing at the same voltage as that at which the dc resistance starts falling rapidly (see Figure 1b), the EMF is presumably caused by a strain dependent CDW conductivity, i.e. $\sigma_{CDW} = \sigma_{CDW}(V-V_T, \varepsilon)$, where $\varepsilon$ = torsional strain.

For example, because the CDW wavevector is strain dependent [4,5], the threshold voltage is also expected to be [1], so $d\sigma_{CDW}/d\varepsilon = \partial\sigma_{CDW}/\partial\varepsilon - (\partial\sigma_{CDW}/\partial V) \, dV_T/d\varepsilon$. If the second term dominates, then, for fixed applied torque, the induced EMF should be proportional to $Z(V) \equiv I_{dc} (J/J_0) \, dR/dV$, i.e. the change in threshold voltage $\Delta V_T = \Delta V_{ac}/Z$. In the inset to Figure 3, we plot the dc voltage dependence of $\Delta V_{ac}/Z$ for the two samples. It is seen that for V > 150 mV this normalized EMF is approximately constant, suggesting that, for voltages well above threshold, the major contribution to the piezoresistance is in fact the strain dependence of the threshold field. Assuming that we are twisting the samples by ~ 5°, our results correspond to relative changes in threshold voltage $\Delta V_T/V_T$ between $2 \times 10^{-4}$ (A) and $10^{-3}$ (B) per degree twist, a few times larger than estimates of relative changes in CDW wavevector [4]. (Note that torsional strain is inhomogeneous and shape dependent, so even for similar twist angles, the two samples will have different distributions of strain.)

In summary, we have studied the voltage and frequency dependence of the voltage-induced torsional strain [6,7] in o-TaS$_3$. The onset of the VITS lies below the threshold for elastic changes, suggesting that it is caused by CDW polarization interacting with extended lattice defects. The strain changes very slowly in response to changing voltage, but the dynamic response appears to be qualitatively very sample dependent. A strain-induced EMF is also observed, but its response is much faster and presumably reflects a piezo-conductance of the sliding CDW, e.g. due to a strain dependent threshold voltage.

We thank V. Ya.Pokrovskii, S.G. Zybstev, W.L. Fuqua, and K.-W. Ng for helpful discussions and R.E. Thorne for providing samples. This research was supported by the National Science Foundation, grants #DMR-0800367, DMR-0400938, and EPS-0814194.

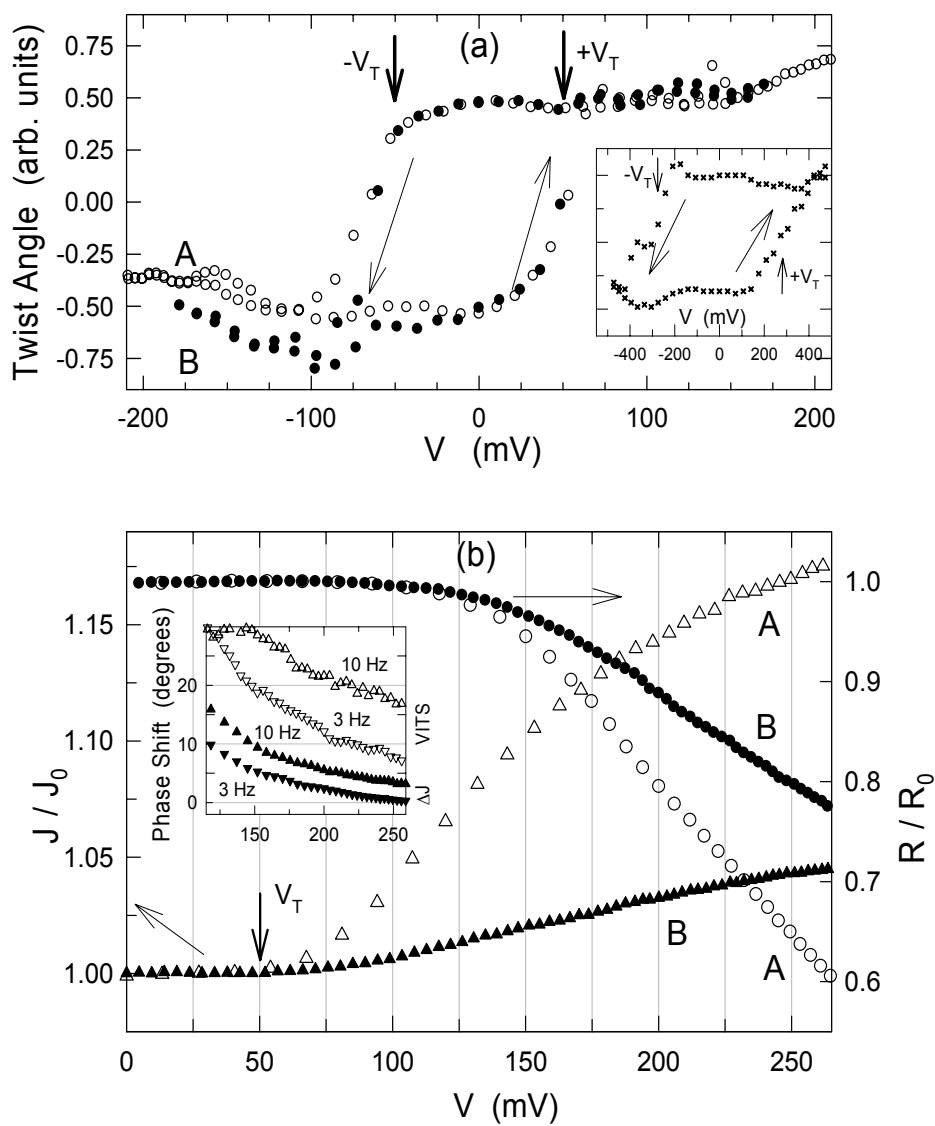

**Figure 1.** a) Voltage induced torsional strain vs. dc voltage for samples A and B. The inset shows the VITS response for a third sample. The vertical arrows show the threshold voltages as determined from the shear compliance. b) The relative changes in resistance (right scale) and the effective 10 Hz shear compliance (left scale) vs. dc voltage for samples A and B. The inset compares the phase shifts at 10 Hz and 3 Hz of the VITS signal (open symbols) and the change in compliance (solid symbols) for sample A.

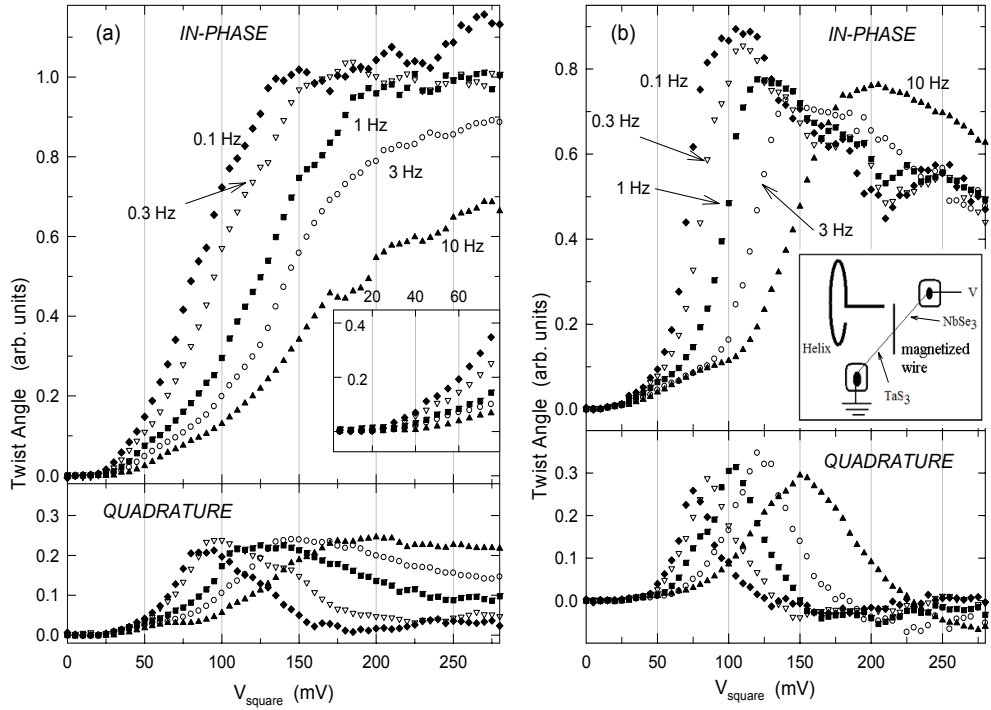

**Figure 2.** Oscillating voltage induced torsional strain as a function of square-wave voltage at several square-wave frequencies for a) sample A and b) sample B; both the strain in-phase (top panels) and in quadrature (bottom panels) are shown. Inset (a): Magnitude of the oscillating strain as function of voltage at small voltages for sample A at the same frequencies as shown in the main panel. Inset (b): Schematic of sample measurement setup. The cavity resonant frequency (~ 430 MHz) depends on the capacitance between the helix tip and magnetized wire.

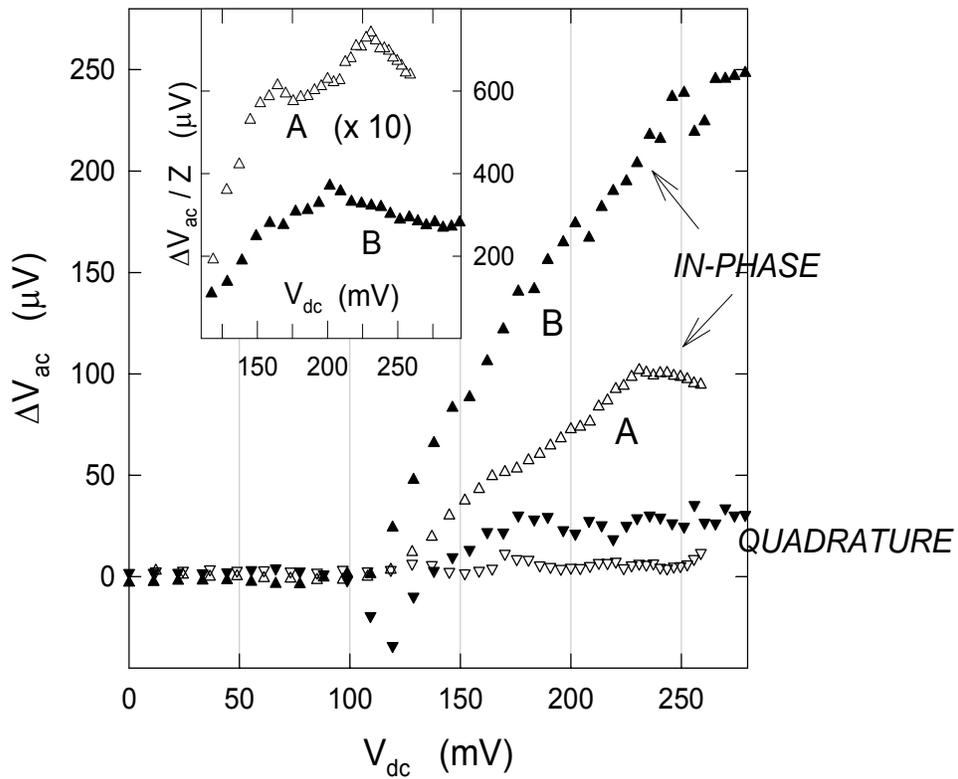

**Figure 3.** AC voltage induced by 10 Hz oscillating strain as function of dc voltage for samples A and B. Inset: Induced ac voltages normalized by the parameter Z(V), defined in text. (Note that the normalized voltages for A are multiplied by 10.)